\documentclass{article}
\usepackage[utf8]{inputenc}
\usepackage{authblk}
\usepackage{setspace}
\usepackage[margin=1.25in]{geometry}
\usepackage{graphicx}
\graphicspath{ {./figures/} }
\usepackage{subcaption}
\usepackage{amsmath}
\usepackage{booktabs}
\usepackage{multirow}
\usepackage{float}
\usepackage{url}
\usepackage{natbib}
\title{Towards NoahMP-AI: Enhancing Land Surface Model Prediction with Deep Learning}

\author[1*$\dag$]{Mahmoud Mbarak}
\author[1]{Manmeet Singh}
\author[1]{Naveen Sudharsan}
\author[1]{Zong-Liang Yang}

\affil[1]{Department of Earth and Planetary Sciences, Jackson School of Geosciences, University of Texas - Austin, USA}
\affil[*]{Corresponding author: mbarak@utexas.edu}

\date{}

\begin{document}

\maketitle

\begin{abstract}
Accurate soil moisture prediction during extreme events remains a critical challenge for earth system modeling, with profound implications for drought monitoring, flood forecasting, and climate adaptation strategies. While land surface models (LSMs) provide physically-based predictions, they exhibit systematic biases during extreme conditions when their parameterizations operate outside calibrated ranges. Here we present NoahMP-AI, a physics-guided deep learning framework that addresses this challenge by leveraging the complete Noah-MP land surface model as a comprehensive physics-based feature generator while using machine learning to correct structural limitations against satellite observations. We employ a 3D U-Net architecture that processes Noah-MP outputs (soil moisture, latent heat flux, and sensible heat flux) to predict SMAP soil moisture across two contrasting extreme events: a prolonged drought (March-September 2022) and Hurricane Beryl (July 2024) over Texas. When comparing NoahMP-AI with NoahMP, our results demonstrate an increase in R² values  from -0.7 to 0.5 during drought conditions, while maintaining physical consistency and spatial coherence. The framework's ability to preserve Noah-MP's physical relationships while learning observation-based corrections represents a significant advance in hybrid earth system modeling. This work establishes both a practical tool for operational forecasting and a benchmark for investigating the optimal integration of physics-based understanding with data-driven learning in environmental prediction systems.
\end{abstract}

\section{Introduction}

The accurate prediction of soil moisture during extreme weather events represents one of the most pressing challenges in contemporary earth system science. Soil moisture mediates water, energy, and carbon exchanges between the land surface and atmosphere, making its accurate representation essential for drought monitoring, flood forecasting, and climate prediction \cite{seneviratne2010investigating,santanello2018land}. However, current modeling capabilities frequently fail during extreme conditions—precisely when accurate predictions are most urgently needed.

Land Surface Models (LSMs) have evolved to incorporate sophisticated representations of hydrological, thermal, and biogeochemical processes \cite{fisher2020perspectives}. The Noah-MP (Multi-Parameterization) model represents a significant advance in this evolution, offering multiple parameterization options for key processes and demonstrating strong performance across diverse environments \cite{niu2011community,yang2011community}. Despite these advances, LSMs including Noah-MP exhibit significant biases during extreme events due to parameterizations optimized for mean conditions, inadequate representation of land-atmosphere feedbacks, and structural limitations in representing scale-dependent processes \cite{dirmeyer2018verification}. The 2012 U.S. drought, which resulted in \$30 billion in economic losses, exemplifies the consequences of these modeling deficiencies \cite{rippey2015us}.

Two categories of extreme events stress LSMs. During droughts, models must represent slow-evolving soil moisture depletion and vegetation stress responses \cite{miralles2019land}. Major LSMs systematically misrepresent soil moisture memory during these conditions. Conversely, extreme precipitation events such as hurricanes challenge LSMs to represent rapid moisture changes and phenomena like the Brown Ocean Effect, where saturated soils sustain tropical cyclone intensity post-landfall \cite{andersen2013quantifying}.

NASA's Soil Moisture Active Passive (SMAP) mission provides global soil moisture observations that reveal systematic biases in LSM predictions \cite{entekhabi2010soil,reichle2017assessment}. While data assimilation approaches can constrain LSM states with these observations, they provide only temporary corrections without addressing underlying model deficiencies \cite{kumar2014assimilation}.Moreover, these corrections can only be applied retrospectively after SMAP observations become available, rather than providing the predictive capability needed before the events occur.

Recent advances in machine learning offer transformative potential for earth system modeling \cite{reichstein2019deep}. However, purely data-driven approaches may violate physical constraints and fail to generalize beyond training distributions. This has motivated physics-informed machine learning approaches that integrate physical understanding with data-driven learning \cite{karniadakis2021physics}. In land surface modeling, such approaches have shown promise for specific applications including parameter estimation and evapotranspiration modeling \cite{tsai2021calibration,zhao2019physics}, yet a systematic framework leveraging the full complexity of state-of-the-art LSMs remains elusive.

Here we present NoahMP-AI, a novel physics-guided deep learning framework that addresses this gap through three key innovations: (1) using the complete Noah-MP land surface model as a comprehensive physics-based feature generator, (2) employing a 3D U-Net architecture for soil moisture prediction, and (3) demonstrating robust performance across both drought and hurricane conditions. This work establishes a generalizable framework for integrating physics-based models with deep learning.

\section{Materials and Methods}

\subsection{Study Design and Experimental Setup}

We designed our study to evaluate the NoahMP-AI framework. Our analysis focuses on the Texas region (25°N-37°N, 94°W-107°W), which provides an ideal test bed due to its diverse climate zones and susceptibility to both drought and hurricane events.

For all experiments, we configured Noah-MP v5.0 \cite{niu2011community,yang2011community} in offline mode at 9 km horizontal resolution to match SMAP observations. The model uses four soil layers (0-10, 10-40, 40-100, 100-200 cm) with hourly NLDAS-2 atmospheric forcing data. 

\textbf{Case Study 1: Extended Drought Conditions (March-September 2022)}\\
The 2022 Texas drought represents a prolonged extreme event characterized by persistent precipitation deficits, extreme temperatures, and progressive soil moisture depletion. Over 70\% of Texas experienced severe to exceptional drought conditions during this period. We generated 684 ensemble members, each covering 2-week simulation periods with perturbed initial soil moisture and temperature states, providing diverse atmospheric forcing sequences for robust training and evaluation.

\textbf{Case Study 2: Hurricane Beryl (July 2024)}\\
Hurricane Beryl made landfall near Matagorda, Texas, on July 8, 2024, as a Category 1 storm with maximum sustained winds of 80 mph. This event tests the framework's ability to capture rapid soil moisture transitions and maintain physical consistency during extreme wetting. We generated 88 ensemble members covering 3-day periods around the hurricane event, with ensemble diversity created through varying initialization times.

For both case studies, Noah-MP provides three key physics-based features that serve as inputs to our deep learning framework: surface soil moisture (0-10 cm, m³/m³), latent heat flux (W/m²), and sensible heat flux (W/m²). 

\begin{table}[h]
\centering
\caption{Summary of experimental design for drought and hurricane case studies}
\label{tab:experiments}
\begin{tabular}{lcc}
\toprule
\textbf{Parameter} & \textbf{Drought Study} & \textbf{Hurricane Study} \\
\midrule
Event period & March-September 2022 & July 2024 \\
Ensemble members & 684 & 88 \\
Simulation length & 2 weeks & 3 days \\
Temporal Resolution & 3-hourly & 3-hourly \\
\bottomrule
\end{tabular}
\end{table}

These features encapsulate the model's representation of water and energy balance processes, providing rich physical information for the deep learning component.

\subsection{SMAP Satellite Observations}

The SMAP Level-4 (L4) surface soil moisture (0-5cm) product serves as our observational target \cite{reichle2017assessment}. SMAP L4 provides global soil moisture estimates at 9 km spatial resolution (EASE-Grid 2.0 projection) and 3-hourly temporal frequency through the assimilation of SMAP brightness temperature observations into the NASA Catchment land surface model. 
Quality flags are used to exclude pixels affected by snow cover, frozen soil conditions, or radio frequency interference. For this study, we bilinearly interpolated SMAP L4 data to the 9 km Noah-MP grid and applied quality control masks to ensure training exclusively on high-confidence observations.

\subsection{NoahMP-AI Architecture: Physics-Guided Deep Learning Framework}

\begin{figure}[h]
    \centering
    \includegraphics[width=0.9\textwidth]{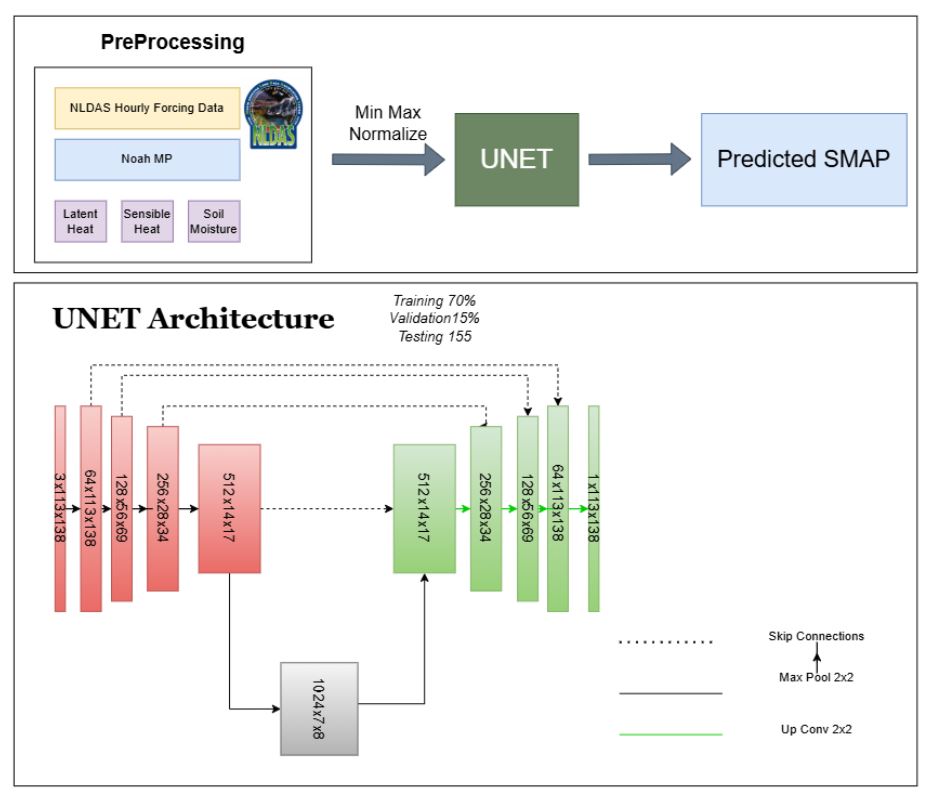}
    \caption{Architecture of the NoahMP-AI framework showing the 3D U-Net encoder-decoder structure.}
    \label{fig:architecture}
\end{figure}

The NoahMP-AI framework employs a 3D U-Net convolutional neural network architecture \cite{ronneberger2015u} specifically adapted for spatio-temporal soil moisture prediction (Figure \ref{fig:architecture}). The network accepts a 4D tensor of shape (batch\_size, channels=3, time, height, width) where the three channels correspond to Noah-MP soil moisture, latent heat flux, and sensible heat flux.

The encoder progressively reduces spatial dimensions through four blocks with increasing feature channels (64, 128, 256, 512), using 3D convolutions, batch normalization, ReLU activations, and max pooling. A bottleneck layer with 1024 channels captures the most abstract representations before the decoder reconstructs the spatial resolution through transposed convolutions and skip connections from corresponding encoder levels. The final 1×1×1 convolution with ReLU activation ensures physically consistent non-negative soil moisture predictions. Skip connections throughout the network preserve Noah-MP's fine-scale physics-based features while the hierarchical structure enables learning of both local corrections and large-scale bias patterns.

\subsection{Training Configuration and Optimization}

\begin{table}[h]
\centering
\caption{Training configuration and hyperparameters for the NoahMP-AI framework}
\label{tab:training_config}
\begin{tabular}{lcc}
\toprule
\textbf{Parameter} & \textbf{Drought Study} & \textbf{Hurricane Study} \\
\midrule
Total samples & 684 ensembles × 112 timesteps & 88 ensembles × 24 timesteps \\
Training set & 70\% (53,625 samples) & 70\% (1,478 samples) \\
Validation set & 15\% (11,490 samples) & 15\% (317 samples) \\
Test set & 15\% (11,490 samples) & 15\% (317 samples) \\
Batch size & 32 & 16 \\
Learning rate & 1e-5 (with decay) & 1e-5 (with decay) \\
Optimizer & AdamW & AdamW \\
Loss function & MAE & MAE \\
Early stopping & 20 epochs patience & 20 epochs patience \\
Maximum epochs & 200 & 200 \\
\bottomrule
\end{tabular}
\end{table}

The training process employs several strategies to ensure robust learning. We use mean absolute error (MAE) as the loss function:
\begin{equation}
\mathcal{L} = \text{MAE}(\hat{y}, y)
\end{equation}
where $\hat{y}$ represents predicted soil moisture and $y$ represents SMAP observations. To ensure independent evaluation, we split data at the ensemble level rather than randomly sampling individual timesteps, preventing data leakage between training and test sets. The AdamW optimizer with learning rate decay provides stable convergence, while early stopping with 20 epochs patience prevents overfitting during the training process.

\section{Results}
\subsection{Drought Case Study}

The NoahMP-AI framework demonstrates significant improvements in soil moisture prediction accuracy during the 2022 Texas drought compared to standalone Noah-MP simulations. The framework achieves substantial error reductions, with mean absolute error decreasing from 0.0642 to 0.0313 m³/m³ and root mean square error improving from 0.0831 to 0.0440 m³/m³ (approx. 50\% reduction). Analysis across all test ensemble members reveals that NoahMP-AI outperformed Noah-MP at every single timestep (112/112), demonstrating consistent enhancement in predictive skill while preserving physically meaningful spatial and temporal patterns.

\begin{figure}[H]
    \centering
    \includegraphics[width=\textwidth]{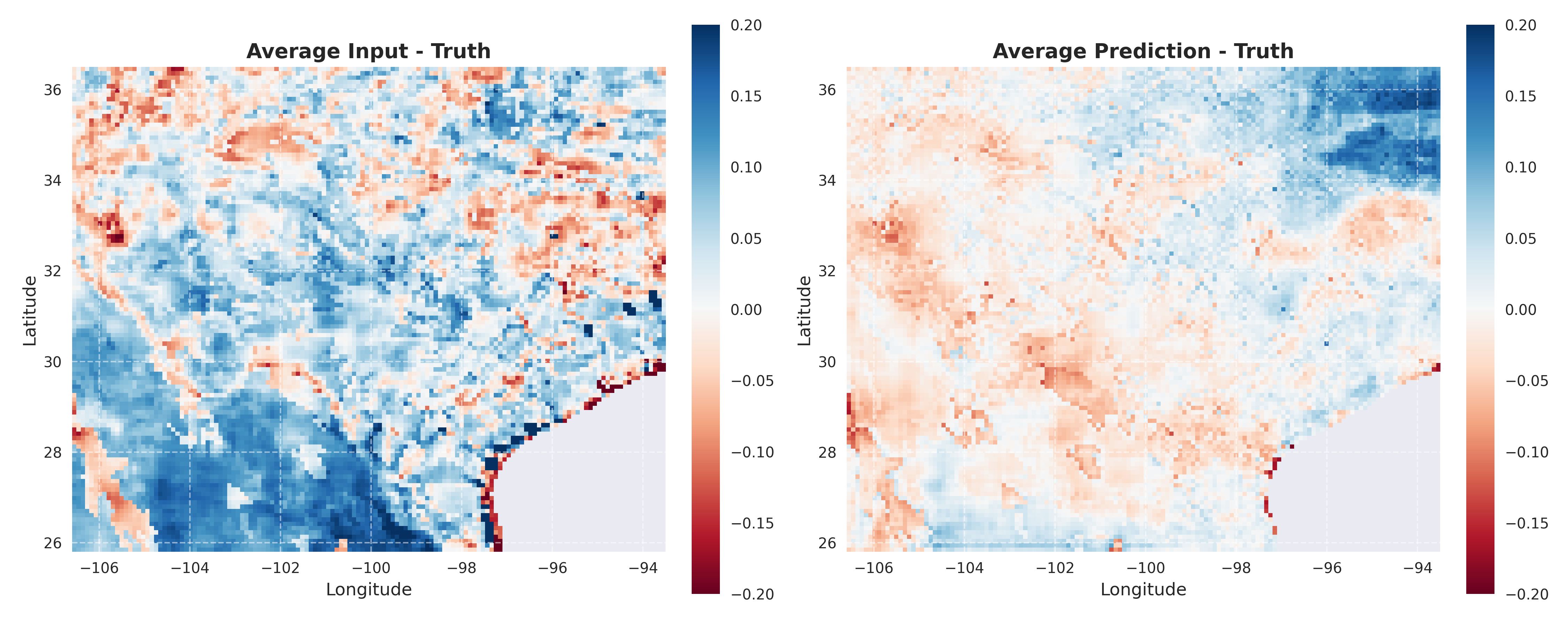}
    \caption{Spatial comparison of average soil moisture bias during the drought period. (a) Noah-MP (b) NoahMP-AI}
    \label{fig:spatial_comparison}
\end{figure}

The spatial bias analysis (Figure \ref{fig:spatial_comparison}) reveals the geographic distribution of model improvements. Noah-MP exhibits pronounced wet biases (blue regions) across southern and central Texas, with scattered dry biases (red regions) throughout the domain. NoahMP-AI effectively reduces these systematic errors, producing bias patterns much closer to zero (white regions) across the majority of the domain while maintaining spatial coherence.

\begin{figure}[H]
    \centering
    \begin{subfigure}[b]{0.48\textwidth}
        \includegraphics[width=\textwidth]{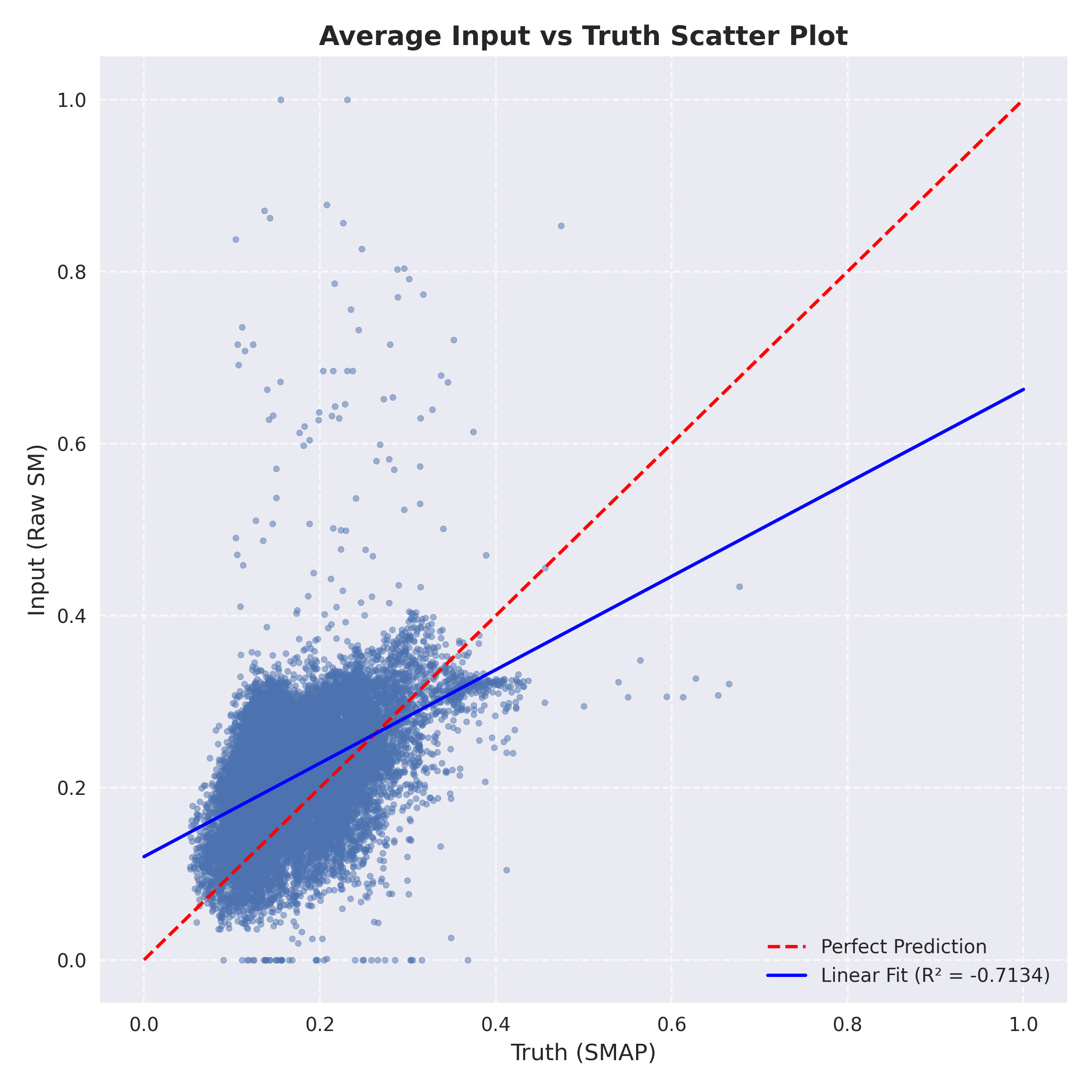}
        \caption{Noah-MP vs SMAP}
    \end{subfigure}
    \hfill
    \begin{subfigure}[b]{0.48\textwidth}
        \includegraphics[width=\textwidth]{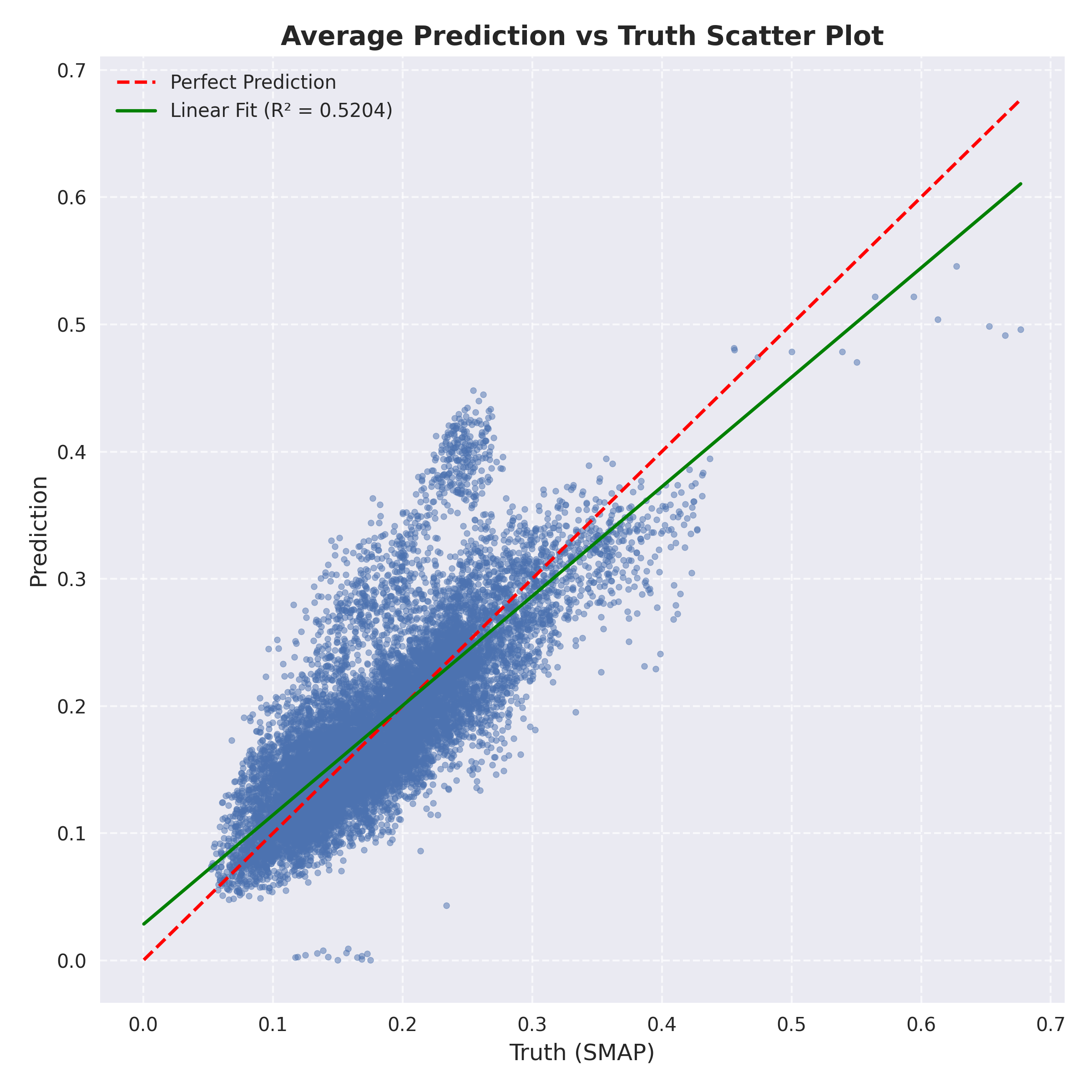}
        \caption{NoahMP-AI vs SMAP}
    \end{subfigure}
    \caption{Scatter plot comparison of soil moisture predictions against SMAP observations. (a) Noah-MP (b) NoahMP-AI}
    \label{fig:scatter_analysis}
\end{figure}

The scatter plot analysis (Figure \ref{fig:scatter_analysis}) demonstrates the fundamental transformation in prediction capability. Noah-MP's negative R² values indicate performance worse than climatological mean, with systematic underestimation particularly severe in mid-to-high moisture ranges. This reflects the model's tendency to over-drain soils during drought conditions, likely due to limitations in representing capillary rise and vegetation stress responses. NoahMP-AI transforms this relationship, achieving positive correlations across the full moisture spectrum with predictions clustering near the 1:1 line.

\begin{figure}[H]
    \centering
    \includegraphics[width=0.8\textwidth]{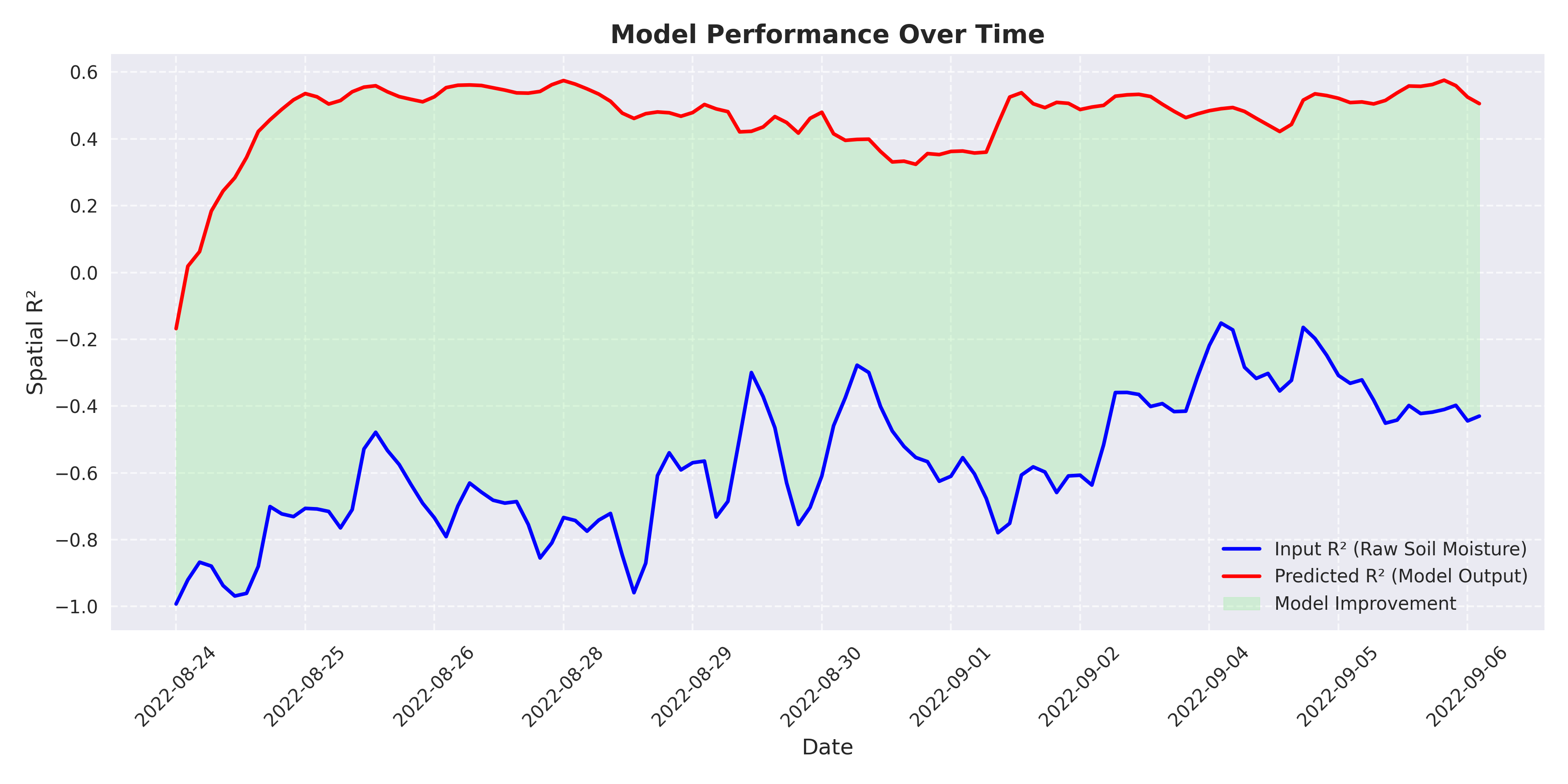}
    \caption{Temporal evolution of spatial R² throughout a representative test ensemble. NoahMP (blue) and NoahMP - AI (Blue)}
    \label{fig:temporal_performance}
\end{figure}

The temporal evolution analysis (Figure \ref{fig:temporal_performance}) confirms the stability of performance gains across varying atmospheric conditions. NoahMP-AI maintains consistently superior performance throughout the simulation period, with the framework demonstrating robust generalization across different drought phases including varying precipitation patterns, temperature extremes, and seasonal transitions. The sustained improvement gap indicates that learned corrections capture fundamental process biases rather than fitting to specific meteorological states.

\subsection{Hurricane Case Study}
The hyrricance case study, involving an independently trained NoahMP-AI represents rapidly-changing soil moisture dynamics captures the distinct bias patterns associated with extreme precipitation events.

\begin{figure}[H]
    \centering
    \includegraphics[width=0.8\textwidth]{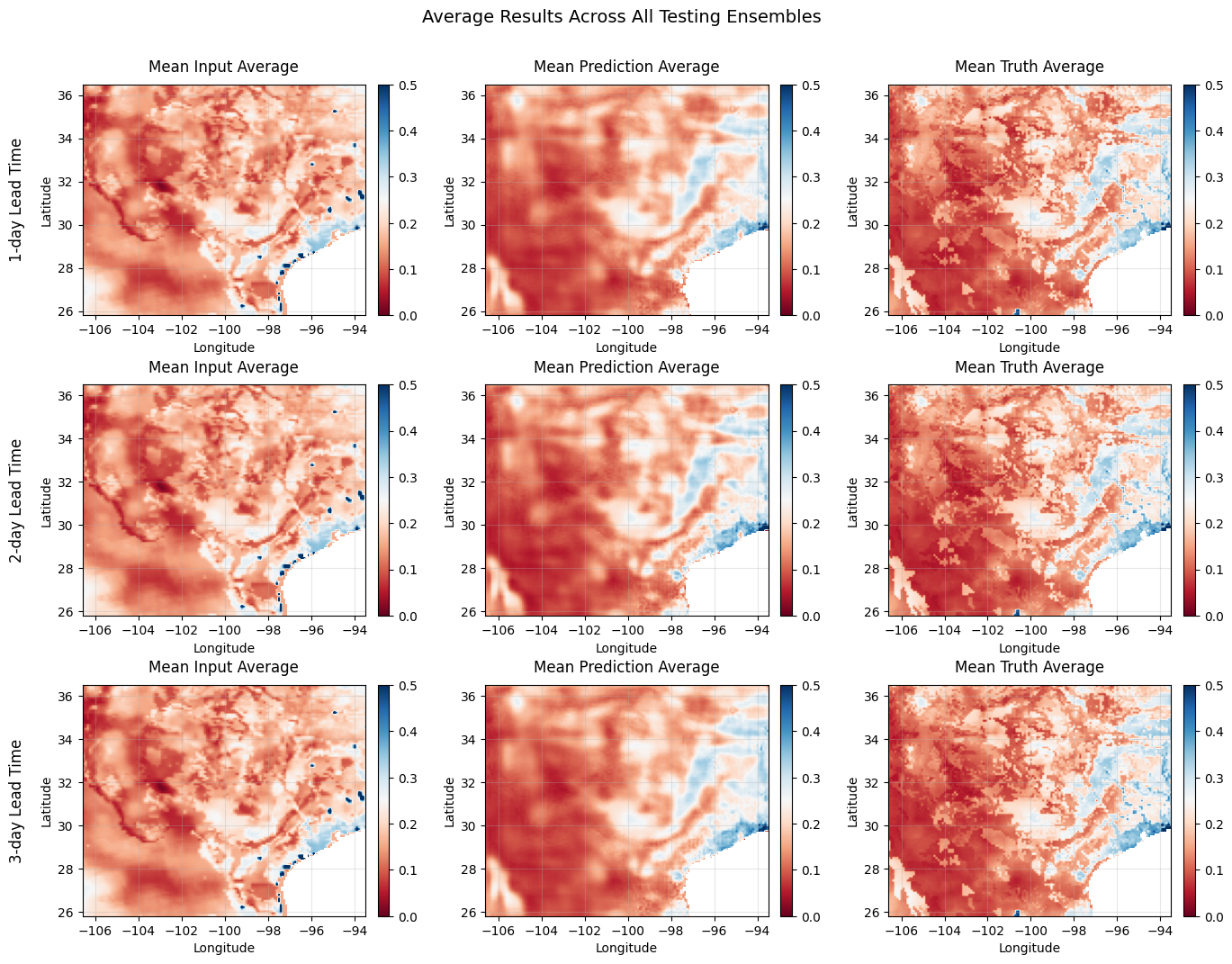}
    \caption{Hurricane Beryl case study: Comparison of soil moisture predictions across different lead times. Each row represents different prediction horizons: 1-day (top), 2-day (middle), and 3-day (bottom) lead time averages. The columns show Noah-MP soil moisture input (left), ML-predicted soil moisture (middle), and SMAP products (right).}
    \label{fig:hurricane_leadtime}
\end{figure}

Figure \ref{fig:hurricane_leadtime} demonstrates the framework's performance across different prediction lead times during Hurricane Beryl. NoahMP-AI model maintains superior performance over Noah-MP across all three lead times, with R² values consistently exceeding 0.75 compared to negative or weakly positive values for the baseline model. The 1-day lead time shows the strongest performance (R² ~ 0.85), while 3-day predictions maintain significant improvement (R² ~ 0.78) despite the increased forecast uncertainty. The framework achieves also a substantial error reductions, with mean absolute error decreasing from 0.0632 to 0.0241 m³/m³ and root mean square error improving from 0.0822 to 0.0348 m³/m³ (approximately 60\% and 58\% reduction, respectively). 

\section{Discussion}

 NoahMP-AI demonstrate the transformative potential of physics-guided deep learning for earth system modeling. By achieving significant R² improvements while maintaining physical consistency, our framework addresses long-standing challenges in LSM performance during extreme events and establishes a new paradigm for integrating comprehensive physics with data-driven learning.

The success of our approach stems from three key design principles. First, using the complete Noah-MP model as a feature generator rather than replacing components preserves the full complexity of physical process representations while enabling targeted bias correction. Second, the 3D U-Net architecture effectively captures both local bias patterns and large-scale systematic errors through its hierarchical structure and skip connections, which preserve fine-scale physics-based features while learning multi-scale corrections \cite{reichstein2019deep,rasp2018deep}. Third, our ensemble-based data splitting ensures rigorous evaluation on completely independent atmospheric forcing sequences, providing confidence for operational deployment.

The framework's consistent performance under various conditons shows its robustness and architectural flexibility. The temporal stability suggests that the learned corrections capture fundamental biases in Noah-MP's physical processes rather than fitting to specific meteorological conditions. This represents a significant advance over traditional bias correction methods that typically rely on statistical post-processing without leveraging physical relationships \cite{wood2008correcting}.

For operational applications, the improved spatial pattern representation has direct implications for drought monitoring, agricultural decision-making, and water resource allocation. Current operational frameworks, including the U.S. Drought Monitor and NOAA's National Water Model, rely heavily on LSM outputs \cite{svoboda2002drought,gochis2015great}. NoahMP-AI's ability to provide bias-corrected estimates in near real-time makes it suitable for integration into these operational workflows, particularly for drought early warning systems where accurate spatial delineation of severity is crucial.

However, several limitations must be acknowledged. The current study focuses on a single geographic region with specific characteristics, and transferability to other regions requires further investigation, though physics-guided design principles should facilitate adaptation through transfer learning. Our evaluation covers specific extreme events, and long-term assessment across multiple cycles would provide stronger evidence for operational reliability. The framework's reliance on SMAP observations introduces dependency on continued satellite data availability, though the architecture could potentially be adapted to alternative observational constraints. Additionally, while we demonstrate improved surface soil moisture prediction, extending to full soil profiles represents an important future direction for agricultural applications.

Our results contribute to the question of optimal integration between physics-based modeling and data-driven learning in Earth system science. The framework essentially learns a high-dimensional transfer function between Noah-MP's process-based predictions and satellite-observed reality, with generalization across diverse atmospheric conditions. Future work should investigate the optimal architectural complexity for effective bias correction, thereby establishing efficiency-accuracy trade-offs for operational implementation. Additionally, extending the framework with interpretability techniques could provide insights into which Noah-MP processes contribute most to prediction errors, potentially informing targeted improvements in physical parameterizations and revealing universal bias patterns across different LSMs. And finally, ensuring full water balance checks at the pixel level to allow for fully coupled LSM/AI paradigm.

\section{Conclusions}

This study presents NoahMP-AI, a physics-guided deep learning framework that substantially enhances soil moisture prediction during extreme events. By integrating the complete Noah-MP land surface model with a 3D U-Net architecture, we achieve dramatic improvements in prediction accuracy while maintaining physical consistency and operational feasibility.

The framework's success across contrasting extreme events—drought and hurricane conditions—validates the effectiveness of comprehensive physics integration over component replacement approaches. This establishes a new paradigm for earth system modeling where complete physical process representations are preserved while enabling targeted data-driven bias correction.

As extreme events become more frequent under climate change, the NoahMP-AI framework represents both a practical tool for operational forecasting and a conceptual advance in optimally integrating physical understanding with machine learning for societal resilience.

\section*{Data and Code Availability}

The Noah-MP model is publicly available at \url{https://github.com/NCAR/noahmp}. SMAP L4 data can be accessed through NASA Earthdata (\url{https://earthdata.nasa.gov/}). NLDAS-2 forcing data is available from NASA GES DISC.

\newpage
\bibliography{references}

\end{document}